\documentclass[floatfix,twocolumn,showpacs,aps,tightenlines,prl]{revtex4}
\usepackage{graphicx}
\usepackage{color}
\usepackage{dcolumn}
\usepackage{bm}

\usepackage[mathscr]{eucal}
\usepackage{latexsym}
\usepackage{stmaryrd}

\usepackage{mathrsfs}
\usepackage{amssymb}
\usepackage{amsmath}
\usepackage{amsfonts}
\usepackage{longtable}
\usepackage{xspace}

\newcommand{\beq}{\begin{equation}}
\newcommand{\eeq}{\end{equation}}
\newcommand{\KMS}{\rm km\,s^{-1}}

\newcommand{\be}{\begin{equation}}
\newcommand{\ee}{\end{equation}}
\newcommand{\bea}{\begin{eqnarray}}
\newcommand{\eea}{\end{eqnarray}}
\newcommand{\bes}{\begin{subequations}}
\newcommand{\ees}{\end{subequations}}

\newcommand{\hn}{\hat{{n}}}

\newcommand{\hangup}{{\it hangup}\xspace}
\newcommand{\ang}[2]{\{#1, #2\}}
\newcommand{\carpet}{{\sc Carpet}\xspace}
\usepackage{bm}

\begin{document}

\title{Where angular momentum goes in a precessing black hole binary}
\author{Carlos O. Lousto}
\author{Yosef Zlochower} 
\affiliation{Center for Computational Relativity and Gravitation,
School of Mathematical Sciences,
Rochester Institute of Technology, 85 Lomb Memorial Drive, Rochester,
 New York 14623}

\date{\today}

\begin{abstract}

We evolve a set of 32 equal-mass black-hole binaries with collinear
spins (with intrinsic spin magnitudes $|\vec{S}_{1,2}/m^2_{1,2}|=0.8$)
to study the effects of precession in the highly nonlinear  plunge and
merger regimes.  We compare the direction of the instantaneous
radiated angular momentum, $\widehat{\delta J}_{\rm rad}(t)$, to the
directions of the total angular momentum, $\hat{J}(t)$, and the
orbital angular momentum, $\hat{L}(t)$.  We find that $\widehat{\delta
J}_{\rm rad}(t)$ approximately follows $\hat{L}$  throughout the
evolution.  During the orbital evolution and merger, we observe that
the angle between $\vec{L}$ and total spin $\vec{S}$ is approximately
conserved to within $1^\circ$, which allows us to propose and test
models for the merger remnant's mass and spin.  For instance, we
verify that the \hangup effect is the dominant effect and largely
explains the observed total energy and angular momentum radiated by
these precessing systems.  We also verify that the total angular
momentum, which significantly decreases in magnitude during the
inspiral, varies in direction by less than $\sim 5^\circ$.  The
maximum  variation in the direction of $\vec J$ occurs when the spins
are nearly antialigned with the orbital angular momentum. Based on our
results, we conjecture that transitional precession, which would lead
to large variations in the direction of $\vec J$, is not possible for
similar-mass binaries and would require a mass ratio
$m_1/m_2\lesssim1/4$.

\end{abstract}

 \maketitle

\noindent
\section{Introduction and motivation}
The 2005 breakthroughs in Numerical Relativity~\cite{Pretorius:2005gq,
Campanelli:2005dd, Baker:2005vv} allowed for accurate investigations
of the orbital motion of merging black-hole binaries (BHBs) in the
highly-nonlinear regime between the slow inspiral (which can be
modeled by post-Newtonian dynamics) and the post-merger phase
(which can be modeled by black-hole perturbation theory).  In
particular it allowed numerical relativists to quantify the effects of
black-hole spin in the last stages of the merger.  This includes 
notable effects such as the
\hangup effect~\cite{Campanelli:2006uy}, which delays or expedites
the merger depending on whether the spins are aligned or
counter-aligned with the orbital angular momentum, and
the generation of recoils as large as
several thousand $\KMS$ when the spins are (at least partially)
antialigned with each other and  have a nontrivial components in 
the orbital plane~\cite{Campanelli:2007cga, Lousto:2011kp}.

Here we investigate the effects of precession of the BHB system
on the evolution of the total angular momentum, $\vec{J}$, the
orbital angular momentum, $\vec{L}$, the total spin,
$\vec{S}$, and the instantaneous radiated angular momentum
$\overrightarrow {\delta J}_{\rm rad}= -d\vec{J}/dt$, during 
the plunge and merger of the binary,
a regime that cannot be accurately modeled within the post-Newtonian
framework.

Radiation of energy and angular momentum causes  the orbit of a binary
to decay, leading to an eventual merger. If the energy is radiated
preferentially in one direction, the remnant will recoil.  Similarly,
determining how much of, and in which directions, the angular momentum
is radiated allows us to accurately model the remnant's spin.  This
is important, for instance, in  modeling the evolution of black-hole spins
by successive mergers in models of structure growth in the early
(high-z) universe \cite{Volonteri:2012yn,AmaroSeoane:2012zu} as well
as for modeling  gravitational waveforms for source identification
\cite{Schmidt:2012rh,Pekowsky:2013ska,Boyle:2013nka} and for predicting
observational consequences in circumbinary disks~\cite{Noble:2012xz}.

Intimately related to the question of 
what is the dominant contribution to the final
spin of merging black holes is the question of whether
{\it transitional} precession
is possible for equal mass (or other mass ratios)
black holes in the strong field regime of
the plunge and merger. 
Transitional precession occurs when a binary
system transitions from a regime where $\vec J$ is dominated by
$\vec L$ to one where $\vec S$ dominates before the merger, 
passing through a regime where  $\vec L\approx-\vec S$
and $\vec J$ 
can undergo dramatic ``flips'' in direction~\cite{Apostolatos94}. 
The complementary regime where $\vec L$ dominates
is called {\it simple} precession and radiation of angular
momentum takes place along $\hat L\approx\hat J$, which leaves the $\vec J$
direction almost unchanged, although its magnitude decreases steadily.

\section{Numerical Simulations}
To single out precession effects, we choose configurations that
have equal-mass and equal-spin (magnitude and direction) BHs. These
configurations~\cite{Campanelli:2006fy} are symmetric under parity
[i.e.\ $(x,y,z)\to(-x,-y,-z)$]  and, consequently, there is no recoil
of the final remnant.
Such configurations maximize $|\vec S| = |\vec S_1 +
\vec S_2|$ 
and hence the precession in the equal-mass case, which can be seen 
from Eq. (2.23a) of Ref.~\cite{Racine:2008kj},
\beq
\frac{d\vec{S}}{dt} =\frac{7}{2r^3}(\vec{L}\times\vec{S})
+ \frac{3}{r^3}(\hn\cdot\vec{S})(\hn\times\vec{S}),\quad[\vec n =
\vec r_1-\vec r_2]. \label{Sconstevol}
\eeq
We also see from Eq.~(3.28c) of Ref.~\cite{Kidder:1995zr}, 
that the equal-mass-aligned-spins configuration maximizes the 
spin-orbit contribution to the radiation of total angular momentum
perpendicular to $\vec J$ itself.

We vary the polar and azimuthal directions of
the initial spins  $(\theta,\phi)$ with respect to $\hat{L}$
in order to ``cover'' the sphere with 32
simulations to find the specific configurations that maximize
either precession or the evolution of the direction of $\vec J(t)$.
The family of BHBs
considered here are equal-mass, equal-spins, $|\vec{S}_{1,2}/m^2_{1,2}|=0.8$, 
BHBs that are further 
characterized by three parameters, the initial
orbital frequency, the polar inclination of the individual BH spins,
and the azimuthal orientation of the spins. 
 For each polar angle $\theta$ we evolve a set
of six azimuthal angles $\phi=0^\circ, 30^\circ,\cdots,150^\circ$
(except for $\theta=0^\circ$ and $\theta=180^\circ$). We
choose initial polar angles equidistant in $\cos \theta$,
$\theta=0^\circ, 48.2^\circ, 70.5^\circ, 90^\circ, 109.5^\circ,
131.8^\circ, 180^\circ$. 
We start from large enough initial 
separations to ensure 4 to 6 orbits prior to the merger.

Our numerical evolutions are based on the 
{\sc
LazEv}~\cite{Zlochower:2005bj} implementation of the moving puncture
approach~\cite{Campanelli:2005dd, Baker:2005vv} with the conformal
function $W$~\cite{Marronetti:2007wz}. Our code
uses the {\sc Cactus}/{\sc EinsteinToolkit}~\cite{cactus_web,
einsteintoolkit} infrastructure and the 
\carpet~\cite{Schnetter-etal-03b} mesh refinement driver.
We use the {\sc TwoPunctures}
thorn~\cite{Ansorg:2004ds} to generate initial
data and use the {\sc AHFinderDirect}
code~\cite{Thornburg2003:AH-finding} to locate apparent horizons.
We measure the horizon mass and spin using the isolated horizon (IH)
algorithm of~\cite{Dreyer02a}.
We use the ``flux-linkages'' formalism~\cite{Winicour_AMGR},
written in terms of $\psi_4$~\cite{Campanelli:1998jv,Lousto:2007mh}, to calculate the radiated
angular momentum $\overrightarrow{\delta J}_{\rm rad}$.

We define the total angular momentum at time $t$ as
$\vec J(t) = \vec J_{\rm ADM} - \int_0^{t} \overrightarrow{\delta J}_{\rm
rad}(\tau) d\tau
$, which measures the angular momentum on the slice up to the
extraction radius (here $R=100M$). 
We then translate the radiative quantities in
time by $R$,
which is an accurate enough measure of the propagation
time for our current purposes.

We used two different
techniques to measure the direction of the orbital angular momentum
First, we use the coordinate trajectory $\vec r(t)$ and
compute $\hat L_{\rm coor} \propto \vec r(t) \times \dot{\vec r}(t)$;
second, we use 
$\vec L(t)=\vec J(t)-\vec S(t)$, where $\vec S(t)$ is measured using
the IH algorithm (its direction is inferred from the zeros of the
approximate Killing field~\cite{Campanelli:2006fy}). 
We use this second definition in our computations although
we verified that our statements below are in agreement with both definitions.
We also verified~\cite{Lousto:2013wta}
that they both follow the general directions of 
Refs.~\cite{Schmidt:2012rh,O'Shaughnessy:2012vm,Boyle:2013nka}, which
define radiation-based measures of preferred (noninertial) asymptotic frames
(see also Fig.~\ref{fig:directions}).

\section{Results}
\begin{figure}
  \includegraphics[width=.9\columnwidth]{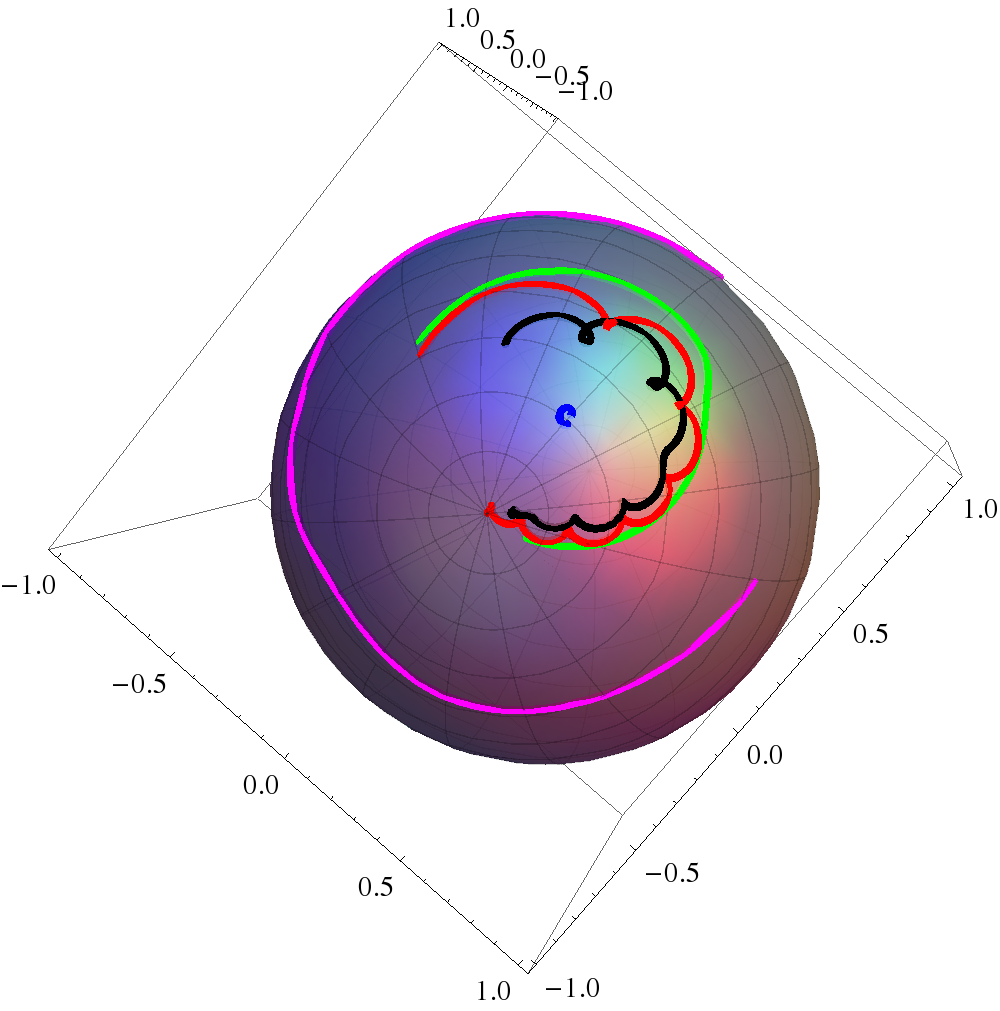}
  \caption{A plot comparing the trajectories of 
(successively outwards curves) $\hat J$ (small blue
curve), $\widehat{\delta J}_{\rm rad}$ (black curve), the coordinate
$\hat L$ (red curve), $\hat L$ as defined by $\vec L = \vec J - \vec
S$ (green curve),  
and $\hat S_1=\hat S_2$ (magenta curve) for a $\theta=90^\circ$ configuration.
All vectors move counterclockwise with time.
 }\label{fig:directions}
\end{figure}

In Fig.~\ref{fig:directions}
we show the results from a prototypical simulation.
We plot the trajectories of $\hat J$,
$\hat L$, $\hat S$, and $\widehat{\delta  J}_{\rm rad}$. 
Note how $\hat J$ is nearly constant (an important feature for modeling),
the large
precession of $\hat L$ and $\hat S$, and that the instantaneous direction of
the radiated angular momentum is close to $\hat L$.

\begin{figure}
  \begin{tabular}{lr}
  \includegraphics[width=0.5\columnwidth]{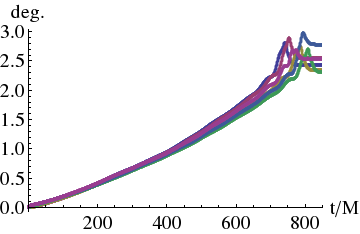} &
  \includegraphics[width=0.5\columnwidth]{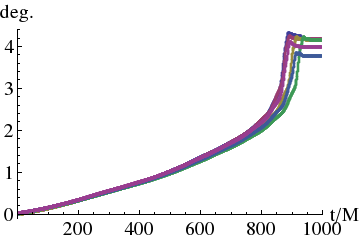}
\\
  \includegraphics[width=0.5\columnwidth]{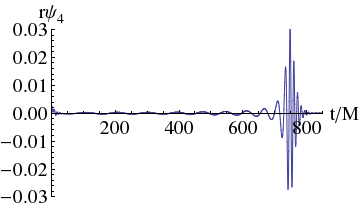} &
  \includegraphics[width=0.5\columnwidth]{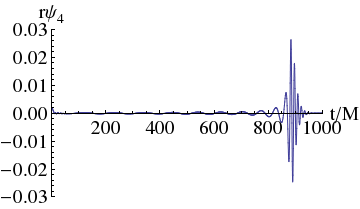}
  \end{tabular}
  \caption{The precession of $\hat J$ (compared to its initial direction)
as a function of time for all the $\theta=110^\circ$  and
$\theta=132^\circ$ configurations (left and
right, respectively).
The bottom plots show the $(\ell=2,m=2)$ mode of $\psi_4$
with the same time scale as a reference. Note that
the precession angle decreases slightly after the large burst of
radiation at merger. All 32 cases studied exhibit {\it simple} precession.
 }\label{fig:jprec}
\end{figure}

We denote the angle between two vectors using
$
\ang{\vec A}{\vec B}= \cos^{-1}(\hat A\cdot \hat B).
$ 
While $\vec J$ changes in magnitude by more than 30\%, the direction
of $\vec J$ is largely unchanged during the entire simulation (a
result also observed in~\cite{Barausse:2009uz}).
 We
thus only observe {\it simple} precession in these equal-mass BHBs. As
seen in Table~\ref{tab:Jprec} and Fig~\ref{fig:jprec}, the net motion
of $\hat J$ (i.e.\ $\ang{\vec J(t)}{\vec J(0)}$) 
is under $5^\circ$. The table also shows how far
$\overrightarrow {\delta J}_{\rm rad}(t)$ deviates from $\hat J$ 
($\ang{\overrightarrow {\delta J}_{\rm rad}(t)}{\vec J(0)}$)
for
selected runs.

\begin{figure}
  \begin{tabular}{lr}
    \includegraphics[width=0.5\columnwidth]{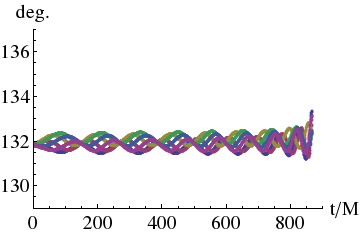} &
    \includegraphics[width=0.5\columnwidth]{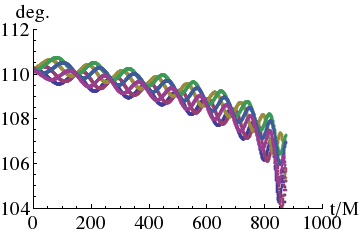}\\
    \includegraphics[width=0.5\columnwidth]{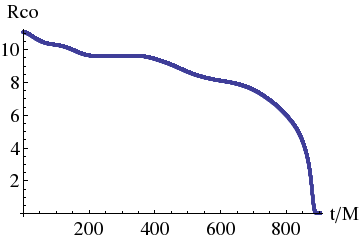} &
    \includegraphics[width=0.5\columnwidth]{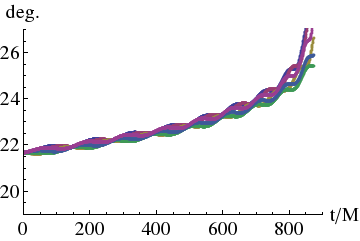}
  \end{tabular}
  \caption{On the left, the upper plot shows the angle between $\hat
L$ and $\hat S_1=\hat S_2$ for all the $\theta=132^\circ$
configurations and the bottom plot shows the orbital separation versus
time for reference.  On the right, the upper plot shows the angle
between $\hat J$  and $\hat S_1=\hat S_2$ and the bottom plot shows
the angle between $\hat J$ and $\hat L$.}
\label{fig:lsprec}
\end{figure}

In Fig.~\ref{fig:lsprec} we show
the time dependence of the angles $\ang{\vec S(t)}{\vec J(t)}$,$\ang{\vec
L(t)}{\vec J(t)} $, and $\ang{\vec S(t)}{\vec L(t)}$.
Compared to the precession of $\vec J(t)$, the angles between 
$\vec J(t)$ and $\vec S(t)$ and $\vec J(t)$
and $\vec L(t)$ vary more strongly, changing by up to $7^\circ$
during the inspiral/plunge. From very large separations, when $\vec J(t)$
is dominated by $\vec L(t)$, the angle between these two vectors secularly
increases due to the radiative loss of angular momentum.
Consequently, the angle between $\vec J(t)$ and $\vec S(t)$ decreases
to keep the angle between $\vec L$ and $\vec S$, 
nearly constant (with oscillations of $\lesssim 1^\circ$
that can be associated with conservative
post-Newtonian dynamics, i.e., nutation). This latter result is very important and will be used 
in the development of an empirical formula for the final remnant mass and spin
since it connects the binary's configuration at large separations with
that at merger~\cite{Lousto:2013wta}.

\begin{table}
  \caption{The change in the direction and magnitude of $\vec J$ for
different $\theta$ configurations. Here $\ang{\vec A}{\vec B} =
\cos^{-1}(\hat A \cdot \hat B)$ measures the angle between two
vectors and $\Delta J=|\vec J(\infty) - \vec J(0)|$ is the magnitude
of the net change in $\vec J$ over the entire simulation. }
\label{tab:Jprec}
  \begin{ruledtabular}
  \begin{tabular}{llllll}
$\cos\theta$   &  2/3 & 1/3 & 0 & -1/3 & -2/3\\
\hline
${\rm max}_{t \phi} \ang{\hat J(t)}{\hat J(0)} $ & $0.86^\circ$ & $1.42^\circ$
& $2.33^\circ$ &$ 3.00^\circ$ &
$4.33^\circ$\\
${\rm max}_ {t \phi} \ang{\widehat{\delta J}(t)}{ \hat J(t)}$ &
$15.68^\circ$ & $21.53^\circ$ & $25.18^\circ$ & $27.39^\circ$ & $27.12^\circ$\\
$|\vec J_{\rm final}|/M^2_{ADM}$   &  0.778 & 0.757 & 0.725 & 0.651 & 0.555\\
$\Delta J/J_{\rm init}$  & 0.352  & 0.321 & 0.290 & 0.311 & 0.332\\
  \end{tabular}
  \end{ruledtabular}
\end{table}

\begin{figure}
  \includegraphics[width=0.49\columnwidth]{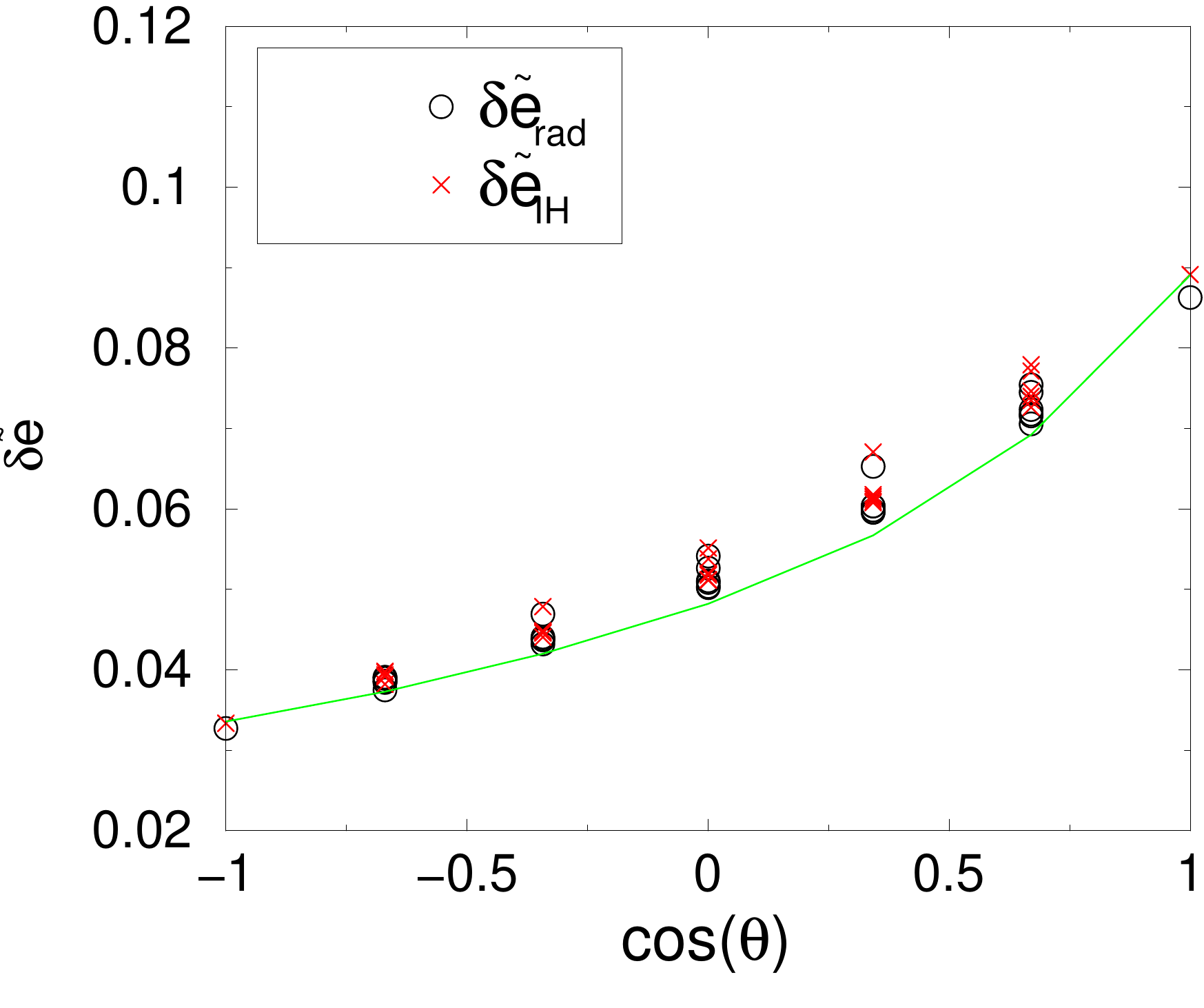}
  \includegraphics[width=0.49\columnwidth]{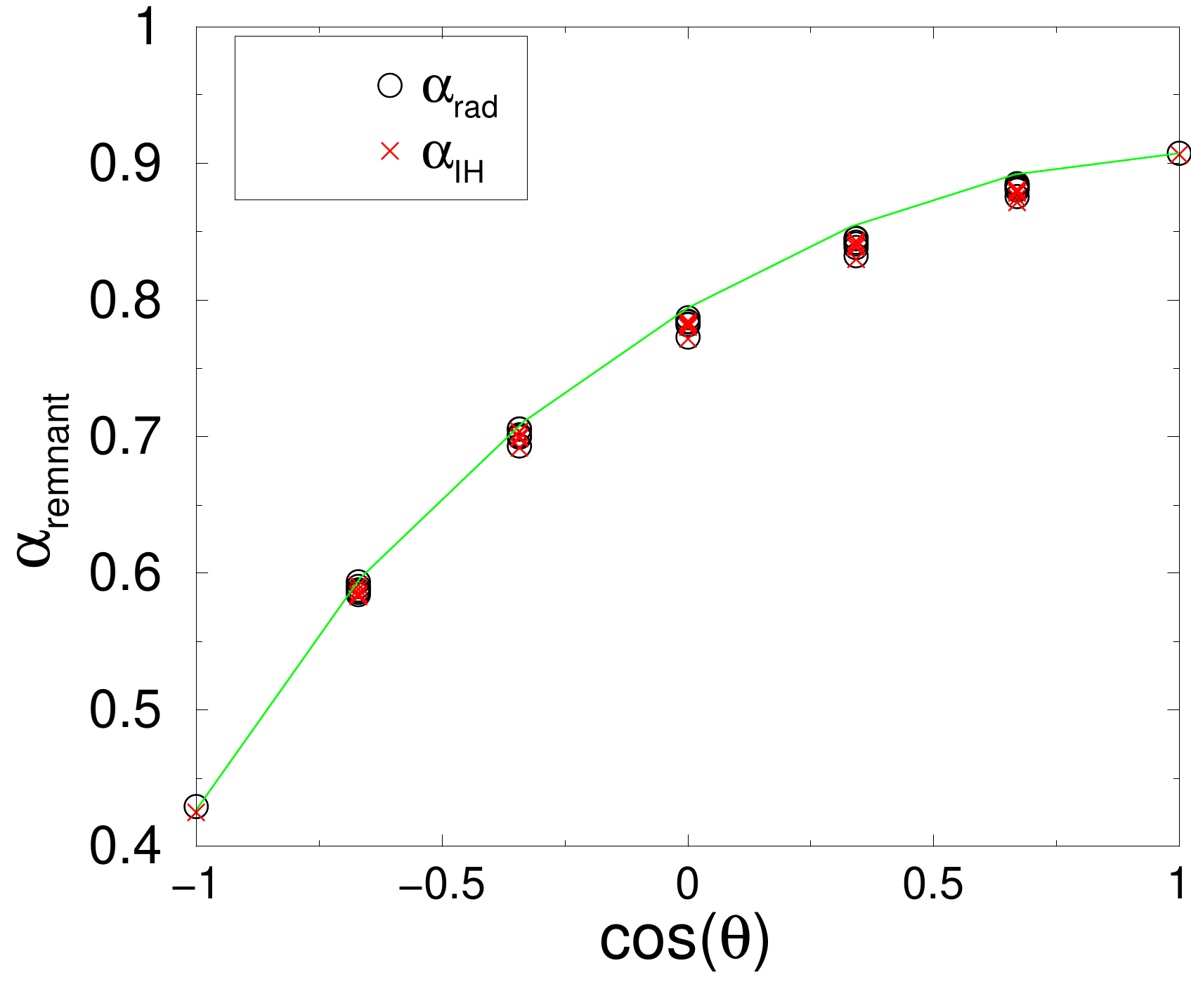}
\caption{Radiated energy and remnant specific spin, as calculated
using the Isolated Horizon formalism and from the radiated
energy-momentum for the 32 simulations, 
as well as the theoretical predictions for these
remnant parameters based on the formulas for the {\it hangup}
remnant from~\cite{Hemberger:2013hsa}.
The different points at the same polar angle
correspond to different azimuthal angles, $\phi=0^\circ-150^\circ$.
No curve fitting was performed in these plots.
}\label{fig:remnant_prediction}
\end{figure}

Here we propose a simple model of the final remnant BH's 
intrinsic spin and mass based on the observations that the total spin
magnitude is conserved to a very high degree during the simulation
and that the direction between the instantaneous orbital angular
momentum and the total spin is approximately conserved
(see also~\cite{Barausse:2009uz}).
Given the assumption that, even for precessing binaries, 
the component of the total spin along the orbital
angular momentum leads to the well-known
\hangup effect~\cite{Campanelli:2006uy}, 
we find that the remnant final spin is given by the \hangup spin with the
addition of the in-plane component of the
initial total spin. That is, the remnant specific spin magnitude is given
by
\begin{equation}
  \alpha_{\rm rem} = \sqrt{F\left(\frac{S_\|}{2 m_h^2}\right)^2  +
S_\perp^2/M_h^2},
\label{eq:alpha}
\end{equation}
where $F(\alpha)$ is the predicted final remnant spin for the
\hangup configuration (an equal-mass, equal-spin, nonprecessing
 configuration where both BH spins are aligned with the orbital
angular momentum and have intrinsic magnitude $\alpha$),
 $S_\|$ is the component of the total spin in the
direction of the orbital angular momentum of the
individual BH spin (the factor of $1/2$ comes from the fact that $\vec
S = 2 \vec S_1= 2 \vec S_2$ here), $S_\perp$ is the magnitude of the total spin
in the orbital plane, $m_h$ is the horizon mass of the individual BHs,
and $M_{h}$ is the mass of the final remnant.
To evaluate $F(\alpha)$ we use the recent work of
Hemberger~{\it et}~al.~\cite{Hemberger:2013hsa}.
We also model
the total radiated energy relative to the mass of the BHB when it was
infinitely separated, i.e.\ $\delta \tilde e = (2 m_h - M_h)/(2 m_h)$.
Here we measure $M_h$ directly using the IH formalism (denoted by
$\delta \tilde
e_{\rm IH}$), and by using the
relation $M_h = M_{\rm ADM} - E_{\rm rad})$ (denoted by $\delta \tilde
e_{\rm rad}$). As shown in Fig.~\ref{fig:remnant_prediction}, we 
find very good agreement between Eq. (\ref{eq:alpha}) and the measured
final spins for all configurations studied here. We also find good qualitative
agreement between the \hangup prediction for the radiated energy
and our results. However, an excess of radiated energy is apparent
due to further nonlinear dependence on the perpendicular component of
the spins to be modeled in \cite{Lousto:2013wta}.

\section{Discussion}
We performed a set of  numerical simulations to explore the 
effects of precession in the fully nonlinear regime.
We chose a simple configuration with equal-mass BHs to complement
particle limit studies, and we chose (relatively) large, collinear 
spins to maximize the precession effects. We found that in this 
highly nonlinear regime the angles between
$\vec{J}$ and $\vec{S}$ or $\vec{L}$ exhibit significant changes on
secular time scales with smaller oscillatory changes on the orbital
time scale (nutation). On the other hand,
the angle between $\vec{S}$ and $\vec{L}$ only shows 
small oscillations on the orbital time scale with no large
secular trend.
While the latter was first observed using low-order post-Newtonian
theory~\cite{Apostolatos94}, we see that, perhaps unexpectedly,
it continues to hold true in the highly-nonlinear merger regime.
We also compute the variation of the direction of the total angular
momentum of the system as seen far from the sources and find that
it varies at most by $\sim5^\circ$ for our runs and hence is limited
by $\sim7^\circ$ for even maximally spinning holes.
These results, connecting early and late time binary dynamics,
are very important for modeling the remnant final masses 
and spins of the merged black hole with applications to cosmological and
astrophysical scenarios, and for gravitational wave modeling.
We verified that these assumptions work well when we compared the total
energy radiated and final spins of our simulation with those of the
\hangup configurations, derived for nonprecessing binaries.

The approximate conservation of the total angular
momentum direction $\hat{J}(t)$
raises the question of where the radiated angular momentum of the system
goes. At large separations, $\vec{J}(t)$ and $\vec{L}$ are almost
aligned with each other, but at the closer separations that we
study here this needs not to be the case. 
As shown in Fig.~\ref{fig:directions},
the radiated angular momentum roughly follows the instantaneous direction
of the orbital angular momentum $\vec{L}(t)$ rather than the more 
constant $\vec{J}(t)$.
Notably, even though a large amount of angular momentum is radiated over time,
this has a relatively small impact on the final direction of $\vec{J}$,
which is a consequence of the fact that $\vec{L}(t)$ itself precesses
roughly around $\vec{J}$, which effectively cancels the net components of
$\overrightarrow{\delta J}_{\rm rad}$ perpendicular to $\vec J(t)$.

Our configurations with equal mass binaries
only exhibit {\it simple} precession; i.e. the direction of
 $\vec{J}(t)$ does not change significantly from its
initial direction. When one considers unequal mass binaries, it is
possible to fine-tune the parameters of the system so that $\vec{J}$
changes
dramatically. To estimate at which mass ratio $q$ this {\it transitional}
precession~\cite{Apostolatos94} may occur, 
we consider the particle limit case of
antialigned, maximally spinning holes 
with the (minimal) orbital angular momentum (at the ISCO~\cite{Misner73}).
Then for $q=m_1/m_2,\ S_1=m_1^2,\ S_2=m_2^2$ we have
\beq
-L^z_{ISCO}=\frac{22}{3\sqrt{3}}\frac{q}{(1+q)^2} m^2\approx
S^z=\frac{(1+q^2)}{(1+q)^2} m^2,
\eeq
which leads to a critical $q_c\approx1/4$, above which transitional precession
is not possible.

The relevance of including the spin dependence in the waveforms of
coalescing black hole binaries to improve the sensitivity of a search, even in the nonprecessing case,
was recently  quantified in Ref.~\cite{Privitera:2013xza}.
Precession effects will further modulate the waveform. Recently, a new
method \cite{Hannam:2013oca, Schmidt:2012rh, Pekowsky:2013ska,
Boyle:2013nka} for generating precessing waveforms by
rotating nonprecessing (hangup) waveforms was developed.
Our work lends support to this method 
since we observe that the binary dynamics are
dominated by the nonprecessing hangup effect. This is a
consequence of the conservation of the angle between $\vec{L}$ and
$\vec{S}$. Note that this hangup waveform will have to be modified during
the latest stages of merger to account for the true total spin as in
Eq.~(\ref{eq:alpha}) (e.g., to get the correct ringdown frequency).
Similarly, an accurate prediction of the final mass and spin of the merged
black hole is crucial for  phenomenological descriptions 
of the gravitational waveforms, such as those based on
SEOBNR~\cite{Taracchini:2012ig}.

In addition, 
the relevance of these results for astrophysical scenarios lies on
the fact that gravitational radiation, even for highly precessing
systems, hardly changes the direction of the total angular momentum 
of the system. Thus suggesting~\cite{Barausse:2009uz},
 for instance, that AGN jets, which are
associated with the direction of the spin of the central black hole engine, 
are good indicators of the direction
of the total angular momentum of the progenitor binary system.
While the accurate modeling of the remnant mass leads to 
specific predictions for the spectrum of the cosmological
gravitational-waves background radiation that could discern
between light and heavy seeds scenarios \cite{Barausse:2012qz}.

\acknowledgments 

The authors gratefully acknowledge the NSF for financial support from Grants
PHY-1305730, PHY-1212426, PHY-1229173,
AST-1028087, PHY-0969855, OCI-0832606, and
DRL-1136221. Computational resources were provided by XSEDE allocation
TG-PHY060027N, and by NewHorizons and BlueSky Clusters 
at Rochester Institute of Technology, which were supported
by NSF grant No. PHY-0722703, DMS-0820923, AST-1028087, and PHY-1229173.

\bibliographystyle{apsrev}
\bibliography{../../Bibtex/references}

\end{document}